\documentclass[12pt,a4paper]{article}
\usepackage[margin=2.3cm]{geometry}
\usepackage{amsmath,amssymb}
\usepackage{graphicx}
\usepackage{siunitx}
\usepackage{booktabs}
\usepackage[colorlinks=true,linkcolor=blue,citecolor=blue,urlcolor=blue]{hyperref}
\usepackage{authblk}
\title{Prediction of the thermal neutron response of grain-boundary impedance in Gd-doped ceria based on the radiation-ionic effect}

\author[1]{Zhongming Zhang}
\affil[1]{School of Engineering, Lancaster University, Lancaster LA1 4YW, United Kingdom\\
\texttt{zhongming.zhang@lancaster.ac.uk}}

\date{\today}

\begin{document}
\maketitle

\begin{abstract}
The radiation-ionic effect was recently reported in \SI{3}{mol\percent} Gd-doped CeO$_2$ (3GDC): under $^{60}$Co gamma irradiation at \SI{35}{Gy/min}, the grain-boundary (GB) ionic resistance of a bulk pellet decreased 680-fold at room temperature, because radiation-generated electrons trapped at the GB core lower the space-charge potential. The effect has so far been demonstrated only with photons. Gd has the largest thermal-neutron capture cross section of any stable element and is at the same time the dopant that creates the GB barrier in GDC, so the same mechanism could link GB impedance to thermal-neutron sensing; whether it does, and under what conditions, has not been examined. Here the thermal-neutron response of 3GDC is modelled with the gamma-calibrated Mott--Schottky/Seager framework, with unmeasured parameters carried as intervals and propagated jointly. Each capture deposits about \SI{50}{keV} by conversion and Auger electrons within \SI{1}{\micro m}, the remaining \SI{8}{MeV} leaving as prompt photons; Gd segregation to the GBs adds a factor of only 1--2 because the electron range exceeds the grain size. A thermal flux of $8\times10^{9}$ (6--12$\times10^{9}$)\,cm$^{-2}$s$^{-1}$ matches the GB-effective generation rate of the \SI{35}{Gy/min} calibration, and the predicted room-temperature GB resistance ratio is 110 (70--170) at $10^{9}$ and 970 (600--1500) at $10^{10}$\,cm$^{-2}$s$^{-1}$, falling to 2 at 61 (55--68) and 81 (73--92)\,$^\circ$C. The neutron case differs from gamma irradiation in three respects: capture events are discrete, so the continuous-source approximation requires the trapped-charge lifetime, which has not been measured; self-shielding (mean free path \SI{0.27}{mm}) makes the response thickness dependent; and capture photons absorbed in the pellet add a gamma term comparable to the local term for pellets of \SI{0.3}{mm} or more. The intervals are conditional on the model and exclude its structural error. The conditions under which the prediction can be tested are given.
\end{abstract}

\section{Introduction}

Grain boundaries (GBs) in doped ceria carry a positive core charge. The space-charge layers on both sides of the boundary are depleted of oxygen vacancies, and the resulting potential barrier $\Delta\Phi$ makes polycrystalline samples orders of magnitude more resistive than single crystals at low dopant content \cite{avila2009dopant,kim2016determining}. Defferriere \emph{et al.} showed that above-bandgap illumination (\SI{375}{nm}) lowers this barrier in \SI{3}{mol\percent} Gd-doped CeO$_2$ (3GDC) thin films \cite{defferriere2022photo}. They then showed that $^{60}$Co gamma irradiation at \SI{35}{Gy/min} lowers the GB resistance of a bulk 3GDC pellet by a factor of 680 at \SI{26.1}{\celsius}, while the bulk (grain-interior) resistance is unchanged within 1\% \cite{defferriere2024ionic}. A closed-form expression relating the resistance ratio to the absorbed photon flux was given in \cite{defferriere2025optoionics}, and the same group has since shown that the GB barrier of bulk 3GDC can be raised or lowered by selective doping of the GB core, with the space-charge contribution identified as the dominant part of the GB resistance \cite{sha2026chemical}. The mechanism is electron trapping at the GB core. The trapped electrons partially neutralize the core charge, and the ionic sublattice then redistributes across the space-charge layer.

To our knowledge, this effect has not been reported by a second group, and it has only been demonstrated with photons. In this work, the response of 3GDC to thermal neutrons is modelled before the experiment is performed. Thermal neutrons are absorbed by $^{157}$Gd (\SI{15.65}{\percent} natural abundance, $\sigma_{\rm th} = \SI{254}{kb}$) and $^{155}$Gd (\SI{14.8}{\percent}, \SI{60.9}{kb}), and each capture releases about \SI{8}{MeV} in a prompt (n,$\gamma$) cascade. Almost all of this energy escapes as MeV photons; about \SI{50}{keV} per capture is deposited within a micrometre of the capture site by internal conversion and Auger electrons. The neutron source therefore differs from gamma irradiation in three respects: it is a set of discrete, micrometre-sized events rather than a continuous uniform generation; it is attenuated within the sample, because Gd itself is the absorber; and it generates its own photon field, so that part of the ``neutron'' response is a gamma response to photons produced in the pellet. The model expresses these differences within the calibrated framework and gives the expected response and the conditions under which it can be measured. The segregation of Gd to GBs could be expected to concentrate the energy at the blocking interfaces; Sec.~\ref{sec:source} shows that it adds a factor of 1--2 at the grain sizes of the published samples. Parameters that are not fixed by nuclear data or by the published measurements are carried as intervals throughout, and every prediction is reported as a median with its 5--95\% range from joint propagation of those intervals. The intervals are conditional on the model and on the assumed (uniform or log-uniform) parameter distributions, which are modelling choices and not measured probability distributions.

This is a modelling study. No new measurements are reported. The two published data sets are used in different roles. The gamma data (bulk pellet, one dose rate, a temperature series) are used for calibration. The thin-film UV data (two temperatures, intensity series) are used as an independent check.

\section{Radiation-ionic framework and calibration}
\label{sec:framework}

\subsection{GB resistance ratio}
In the Mott--Schottky (frozen-dopant) description, the ratio of dark to irradiated GB resistance is \cite{defferriere2025optoionics,vollman1994grain}
\begin{equation}
\frac{R_{\rm gb,D}}{R_{\rm gb,L}} = \sqrt{\frac{\Delta\Phi_{\rm L}}{\Delta\Phi_{\rm D}}}\;
\exp\!\left[\frac{z e\,(\Delta\Phi_{\rm D}-\Delta\Phi_{\rm L})}{k_{\rm B}T}\right],\qquad z = 2,
\label{eq:VW}
\end{equation}
where $\Delta\Phi$ is in volts and $e\Delta\Phi$ is the corresponding barrier energy. The expression is valid for $z e\Delta\Phi_{\rm L}/k_{\rm B}T\gtrsim1$, and it is used only in that range; where a prediction would require $\Delta\Phi_{\rm L}$ below $k_{\rm B}T/ze$, no value is given. The space-charge model has been confirmed for bulk 3GDC by the agreement, within 20\%, between barrier heights obtained from the resistance ratio and from the bias dependence of the GB current, with structural contributions to the GB resistance identified as minor \cite{sha2026chemical}. Independently of Eq.~\eqref{eq:VW}, there is a practical ceiling on the measurable ratio. When the barrier has collapsed, the GB region has bulk-like resistivity and its resistance is of order $R_{\rm bulk}\cdot2\lambda/d_g$, far below $R_{\rm bulk}$; the GB arc then merges with the bulk arc and only the total resistance $R_{\rm gb}+R_{\rm bulk}$ can be measured. The measurable ratio therefore saturates at $(R_{\rm gb,D}+R_{\rm bulk})/R_{\rm bulk}\approx R_{\rm gb,D}/R_{\rm bulk}$, which is a limit on the observable rather than on the GB resistance itself. The gamma experiment gives $R_{\rm gb} = \SI{3.35e7}{\ohm}$ and $R_{\rm bulk} = \SI{1.51e5}{\ohm}$ at \SI{141.7}{\celsius} with activation energies of 1.07 and \SI{0.70}{eV} \cite{defferriere2024ionic}, i.e.\ a ceiling of 222 there and $\SI{1.3e4}{}$ at \SI{25}{\celsius} for the \SI{3}{mol\percent}, \SI{400}{nm}-grain pellet; the measured 680 and the authors' extrapolated 1300 at \SI{21}{\celsius} sit a factor 10--20 below it. All predictions below are capped at this ceiling. For other Gd contents a second anchor is available: for \SI{20}{mol\percent} GDC with \SI{0.75}{\micro m} grains, Bowman \emph{et al.} measured a specific GB conductivity $5\times10^4$ times below the grain-interior value at \SI{200}{\celsius}, with GB and bulk activation energies of 0.92 and \SI{0.87}{eV} \cite{bowman2015electrical}; with a \SI{2.2}{nm} boundary width this corresponds to $R_{\rm gb}/R_{\rm bulk}\approx150$ at \SI{200}{\celsius} and $\sim$310 at \SI{25}{\celsius}. Between the two anchors the ceiling is interpolated log-linearly in $x$; it is a weak function of composition compared with the exponential of Eq.~\eqref{eq:VW}, and the composition dependence in Sec.~\ref{sec:gd} is set mainly by $\Delta\Phi_{\rm D}(x)$ through Eq.~\eqref{eq:VW} itself.

\subsection{Space-charge potential under irradiation}
The steady-state potential follows from the Seager balance of electron capture at GB-core traps against hole recombination \cite{seager1981grain,joshi1990theory}; in the asymptotic form derived in the Supplementary Information of \cite{defferriere2022photo},
\begin{equation}
\Delta\Phi_{\rm L} = \Delta\Phi_{\rm D} - \frac{k_{\rm B}T}{2e}\ln\!\left(1 + \frac{n}{n_{\rm crit}}\right),
\label{eq:seager}
\end{equation}
with $n = G\tau_e$ the free-electron concentration for a continuous source of pair-generation rate $G$ and electron lifetime $\tau_e$, and $n_{\rm crit}=(\sigma_p v_h n_{t,\rm eq})/(\sigma_t v_e N_t)$ the trap parameter. Only $n_{\rm crit}/\tau_e$ is constrained by resistance data; $n_{\rm crit}$ is quoted for $\tau_e = 10^{-7}$\,s but no prediction depends on the split.

\subsection{Calibration on the gamma data}
With the Klein relation $G_\gamma = \dot D\rho/W_{\rm pair}$ and $W_{\rm pair} = 2.8E_g+0.5 = \SI{8.9}{eV}$ \cite{klein1968bandgap}, \SI{35}{Gy/min} gives $G_\gamma = \SI{2.95e15}{cm^{-3}s^{-1}}$. Requiring Eqs.~\eqref{eq:VW}--\eqref{eq:seager} to give 680 at \SI{26.1}{\celsius} with $\Delta\Phi_{\rm D} = \SI{0.20}{V}$ (the value the authors extract from their dark Arrhenius data) yields $n_{\rm crit}/\tau_e = \SI{3.3e12}{cm^{-3}s^{-1}}$ and $\Delta\Phi_{\rm L} = \SI{0.11}{V}$. Since the authors quote the ratio as ``$\approx10^3$'' and extrapolate 1300 at \SI{21}{\celsius}, the calibration ratio is carried as the interval 500--900, which gives $n_{\rm crit}/\tau_e$ = 2.4--\SI{4.5e12}{cm^{-3}s^{-1}}. The calibration temperature is $T_{\rm cal} = \SI{299.25}{K}$ (\SI{26.1}{\celsius}); room-temperature predictions are evaluated at \SI{298.15}{K}, the \SI{1.1}{K} difference being carried through Eq.~\eqref{eq:ncT}.

The gamma experiment also provides a temperature series at the same dose rate \cite{defferriere2024ionic}: above $\sim$\SI{100}{\celsius} the irradiated and dark GB resistances coincide, while below it the irradiated GB resistance has an apparent activation energy of only \SI{0.08}{eV} against \SI{1.07}{eV} in the dark. Reconstructing the ratio from these two activation energies and the \SI{26.1}{\celsius} value, and inverting Eqs.~\eqref{eq:VW}--\eqref{eq:seager} at each temperature, gives $n_{\rm crit}/\tau_e$ rising from \SI{3.3e12}{} at \SI{26}{\celsius} to \SI{3e15}{} at \SI{80}{\celsius}, an Arrhenius slope of $E_{n} = \SI{1.13}{eV}$ (Fig.~\ref{fig:T}a). This slope is essentially the difference of the two stated activation energies and is therefore not an independent measurement. However, it has a physical interpretation. $n_{\rm crit}$ is proportional to the rate at which holes reach the core and recombine with trapped electrons, and the hole diffusivity in the original authors' numerical model is activated with \SI{1.08}{eV} (Supplementary Table~S4 of \cite{defferriere2022photo}). The calibrated temperature dependence is therefore taken as
\begin{equation}
n_{\rm crit}(T)/\tau_e = \left(\frac{n_{\rm crit}}{\tau_e}\right)_{\rm RT}\exp\!\left[\frac{E_n}{k_{\rm B}}\Big(\frac{1}{T_{\rm cal}}-\frac{1}{T}\Big)\right],\qquad E_n = 1.0\text{--}\SI{1.3}{eV},
\label{eq:ncT}
\end{equation}
where the interval for $E_n$ covers the reconstructed slope (1.13), the hole-transport value (1.08) and the uncertainty of the two activation energies from which the slope is derived. The independent check is in Sec.~\ref{sec:transfer}.

\begin{table}[t]
\centering
\caption{Model parameters. A: literature values with small uncertainty. B: literature estimates with stated range, propagated in Sec.~\ref{sec:uncert}. C: calibrated on \cite{defferriere2024ionic}. D: bounded only.}
\label{tab:params}
\small
\begin{tabular}{@{}l l c p{5.0cm}@{}}
\toprule
Parameter & Value & Class & Source \\
\midrule
$\varepsilon_r$ & 37 & A & from bulk capacitance, \cite{defferriere2024ionic} \\
$N_{\rm Ce}$ & \SI{2.51e22}{cm^{-3}} & A & fluorite, $a = \SI{5.412}{\angstrom}$ \cite{defferriere2024ionic} \\
$^{157}$Gd: $\sigma$, $f_{\rm ab}$, $Q$ & \SI{254}{kb}, 0.1565, \SI{7.94}{MeV} & A & ENDF, IUPAC, ENSDF \cite{HELMER2004325} \\
$^{155}$Gd: $\sigma$, $f_{\rm ab}$, $Q$ & \SI{60.9}{kb}, 0.148, \SI{8.54}{MeV} & A & ENDF, IUPAC \\
$E_g$ & \SI{3.0}{eV} & A & \cite{defferriere2024ionic} \\
$\Delta\Phi_{\rm D}$(3\%) & \SI{0.20}{V} & A & \cite{defferriere2024ionic} \\
Calibration point & \SI{26.1}{\celsius}, \SI{35}{Gy/min}, ratio 680 & A & \cite{defferriere2024ionic} \\
$E_a$(GB), dark / irradiated & 1.07 / \SI{0.08}{eV} & A & \cite{defferriere2024ionic} \\
Pellet thickness, grain, density & \SI{0.8}{mm}, \SI{400}{nm}, 92\% & A & \cite{defferriere2024ionic} \\
Ceiling $R_{\rm gb}/R_{\rm bulk}$ (3 / 20\%) & \SI{1.3e4}{} / 250 at RT & A & \cite{defferriere2024ionic} / \cite{bowman2015electrical} \\
\midrule
$W_{\rm pair}$ & 7.6--\SI{10.2}{eV} & B & Klein relation $\pm15\%$ \cite{klein1968bandgap} \\
Local energy per $^{157}$Gd capture & 49--\SI{75}{keV} & B & IC electrons 0.69/capture $\times$ \SI{66.5}{keV}, Auger 4--\SI{7}{keV} \cite{goorley2000electron,HELMER2004325} \\
Local energy per $^{155}$Gd capture & 25--\SI{75}{keV} & D & not tabulated; same order as $^{157}$Gd or half \\
$R_{\rm loc}$ & 0.5--\SI{2}{\micro m} & B & 70\,keV electron CSDA range in CeO$_2$ \cite{berger2017nist} \\
$f_{\rm enh}$ & 1--2 & B & lower end from Eq.~\eqref{eq:fenh}; upper end allows diffusion-limited collection (Sec.~\ref{sec:source}) \\
$\Delta\Phi_{\rm D}(x)$ & $0.20\,(3/x)^{m}$\,V, $m$ = 0.45--0.65 & B & fit to \cite{avila2009dopant}; extrapolated outside 3\% \\
\midrule
$n_{\rm crit}/\tau_e$ (RT) & 2.4--\SI{4.5e12}{cm^{-3}s^{-1}} & C & this work, from the calibration-ratio interval 500--900 \\
$E_n$ & 1.0--\SI{1.3}{eV} & C & this work, Eq.~\eqref{eq:ncT}; central value 1.13 \\
$\mu_{\rm en}/\rho$ (capture photons) & 0.019--\SI{0.029}{cm^2/g} & B & photon energies 1--3 MeV \\
\midrule
$\tau_{\rm rec}$ (RT) & 1--\SI{60}{s} & D & recovery ``within seconds'', equilibration $\sim$\SI{1}{min} \cite{defferriere2024ionic} \\
$\tau_e$ & $10^{-7}$\,s & D & degenerate with $n_{\rm crit}$ \\
\bottomrule
\end{tabular}
\end{table}

\subsection{Time constants}
Two times enter the neutron problem: the time for the impedance to follow a change in trapped charge (ionic redistribution across the space-charge layer, $\tau_{\rm ion}$) and the lifetime of the trapped charge once generation stops ($\tau_{\rm rec}$). The gamma experiment reports that the impedance reached its irradiated value within the $\sim$\SI{1}{min} equilibration allowed, and that it returned to its dark value ``within less than seconds'' after removal from the irradiator at room temperature \cite{defferriere2024ionic}. The recovery bounds both times. The trapped charge disappears, and the ions respond, within seconds. $\tau_{\rm rec}$ at room temperature is therefore taken in the range 1--\SI{60}{s} (the upper end allowing for the equilibration time) and $\tau_{\rm ion}\lesssim\tau_{\rm rec}$. Extrapolating the thin-film photocurrent spectroscopy (\SI{1}{kHz} ionic process at \SI{430}{\celsius}, \SI{0.7}{eV}) to room temperature would give $\tau_{\rm ion}\sim10^3$\,s. This differs from the bulk observation by three orders of magnitude. The bulk value is used here because it was measured on the same type of sample. The reason for the discrepancy (field-assisted drift or a sample difference) is not known.

\section{Thermal-neutron source term}
\label{sec:source}

Natural Gd absorbs thermal neutrons through two isotopes. Per Gd atom the effective cross section is $\sigma_{\rm eff} = f_{157}\sigma_{157}+f_{155}\sigma_{155} = \SI{48.8}{kb}$, of which $^{155}$Gd contributes 18\%. The volumetric capture rate is $R_{\rm cap} = \Phi_n\sigma_{\rm eff}x_{\rm Gd}N_{\rm Ce}$; at \SI{3}{mol\percent}, $R_{\rm cap} = 36.7\,\Phi_n$\,cm$^{-3}$s$^{-1}$ and the macroscopic absorption cross section is $\Sigma_a = \SI{36.7}{cm^{-1}}$.

The $^{157}$Gd capture products have been characterized in detail for gadolinium neutron capture therapy \cite{goorley2000electron,HELMER2004325}. Of the \SI{7.94}{MeV} released, about \SI{7.85}{MeV} leaves as prompt photons with energies up to the full cascade energy and mean free paths of centimetres in CeO$_2$. Per capture, 0.69 internal-conversion electrons of mean energy \SI{66.5}{keV} (discrete lines from 29 to \SI{246}{keV}, dominated by conversion of the 79.5 and \SI{182}{keV} transitions) carry about \SI{46}{keV}; a cascade of 4--10 Auger electrons carries 4--\SI{7}{keV}; and 0.84 Gd X-rays of mean energy \SI{12.8}{keV} carry about \SI{11}{keV}. The electrons stop within $R_{\rm loc}\approx\SI{1}{\micro m}$ in CeO$_2$ \cite{berger2017nist}; the Gd K X-rays (\SI{43}{keV}) travel about \SI{0.1}{mm}. The local energy per $^{157}$Gd capture is therefore carried as 49--\SI{75}{keV}, the range of the tabulations; for $^{155}$Gd, whose conversion-electron spectrum has not been tabulated with the same detail, 25--\SI{75}{keV} is used. With $W_{\rm pair}$ = 7.6--\SI{10.2}{eV} this gives $N_{\rm pair}$ = 5300--8400 pairs per capture (median 6800) in $V_{\rm loc} = \tfrac43\pi R_{\rm loc}^3$, $R_{\rm loc}$ = 0.5--\SI{2}{\micro m}. $V_{\rm loc}$ contains of the order of 100 grains of the \SI{400}{nm} microstructure of \cite{defferriere2024ionic}.

Gd is enriched by a factor $e = 2$--3 within a core region of width $w\approx\SI{2}{nm}$ at the GBs of sintered ceria \cite{bowman2015electrical}. Whether this segregation increases the generation rate seen by the GBs depends on where the carriers of a capture end up, and this is governed by two length scales. The conversion electrons travel $R_{\rm loc}\approx\SI{1}{\micro m}$, far more than the space-charge half-width $\lambda\approx\SI{1}{nm}$ and the grain size $d_g = \SI{400}{nm}$, and they lose only about \SI{60}{eV} in the first \SI{2}{nm} of their track; the thermalized carriers they create are small polarons that move only nanometres within their lifetime (Supplementary Section~17 of \cite{defferriere2022photo}), so they are trapped where they are generated. The fraction of these carriers that falls inside the space-charge regions is therefore the geometric volume fraction of those regions, $2\lambda\cdot3/d_g\approx1.5\%$, whether the capture occurred at a GB or in a grain interior. The Auger electrons are different: with a few keV they stop within $R_A\approx20$--\SI{50}{nm}, so a capture in the GB core deposits a fraction $\approx(\lambda/R_A)[1+\ln(R_A/\lambda)]\approx0.15$ of its Auger energy inside the space-charge slab. Segregation redistributes Gd without changing the average Gd content, so the volumetric capture rate is unchanged; it raises the fraction of captures that occur in the core from its value for a uniform Gd distribution, $p_0 = 3w/d_g\approx1.5\%$, to $e\,p_0$, and lowers the fraction in the grain interior correspondingly. For a uniform distribution the GB-region generation averaged over all capture positions is the geometric value $g_0 = (2\lambda\cdot3/d_g)(N_{\rm IC}+N_A)$, and this average already contains the above-average contribution of core captures, $g_{\rm core}$, and the below-average contribution of interior captures, $g_{\rm int}$, with $(1-p_0)g_{\rm int}+p_0g_{\rm core} = g_0$. With segregation the weighting changes to $(1-ep_0)g_{\rm int}+ep_0g_{\rm core}$, so the enhancement relative to the uniform distribution is
\begin{equation}
f_{\rm enh} = 1 + (e-1)\,p_0\,\frac{g_{\rm core}-g_{\rm int}}{g_0},\qquad
g_{\rm core} \approx g_0 + N_A\,\frac{\lambda}{R_A}\Big[1+\ln\frac{R_A}{\lambda}\Big] + N_{\rm IC}^{\rm slab},
\label{eq:fenh}
\end{equation}
which returns $f_{\rm enh} = 1$ for $e = 1$. The GB core charge used here, $\sim\SI{1.6e14}{}$ elementary charges per cm$^2$, agrees with the net GB charge of \SI{0.26}{C/m^2} obtained for bulk 3GDC from bias-dependent impedance \cite{sha2026chemical}. Here $N_{\rm IC}$ and $N_A$ are the pairs from the conversion and Auger electrons and $N_{\rm IC}^{\rm slab}$ the few pairs from the first \SI{2}{nm} of the conversion-electron track. With $e = 3$, $w = \SI{2}{nm}$, $\lambda = \SI{1}{nm}$ and $R_A = \SI{30}{nm}$, Eq.~\eqref{eq:fenh} gives $f_{\rm enh}\approx1.03$. This is the lower end of what segregation can contribute. A larger enhancement, up to the order of $e$ itself, arises if carrier collection at the boundaries is diffusion limited, i.e.\ if a fraction of the carriers generated in grain interiors recombines before reaching a boundary while those generated at or near a boundary do not; the carrier diffusion length in GDC has not been measured, so this case cannot be excluded for \SI{400}{nm} grains. The enhancement is therefore carried as the interval $f_{\rm enh}$ = 1--2. A deliberate increase of the Gd concentration at the GB core is not a route to a larger response: infiltrating an additional \SI{3}{mol\percent} Gd at the GBs of 3GDC lowers the barrier from 0.27 to \SI{0.19}{V} and removes the non-linear space-charge response altogether \cite{sha2026chemical}, so the capture rate at the boundary and the barrier that is to be modulated cannot be raised independently by Gd alone. The GB-effective generation rate is
\begin{equation}
G_{\rm eff,GB}(\Phi_n) = R_{\rm cap}\,N_{\rm pair}\,f_{\rm enh},\qquad f_{\rm enh}\in[1,2].
\label{eq:Geff}
\end{equation}
At \SI{3}{mol\percent}, a thermal flux of $\Phi_n$ = 8 (6--12)$\times10^{9}$\,cm$^{-2}$s$^{-1}$ delivers the same GB-effective generation rate as the \SI{35}{Gy/min} calibration, while depositing locally only about \SI{25}{Gy/min}. One further effect of a capture in the core should be noted: the conversion and Auger cascade leaves the emitting Gd ion transiently in a high positive charge state at the GB core, of the sign opposite to the trapped electrons; it neutralizes within picoseconds by capturing electrons from its neighbours and involves of the order of ten charges per capture, against the thousands of pairs generated, so it is neglected. The X-rays and the fraction of the cascade photons absorbed in a finite pellet add a uniform generation term that is treated in Sec.~\ref{sec:sample}; it is comparable to or larger than the local term for pellets of practical thickness.

\section{Temporal structure of the source}
\label{sec:regimes}

Equation~\eqref{eq:Geff} treats the source as continuous, as for gamma irradiation. However, a capture event is a point in space and time. The continuous description is valid only if the trapped charge produced by one event is still present when the next event reaches the same volume. The overlap parameter is
\begin{equation}
\eta = R_{\rm cap}\,V_{\rm loc}\,\tau_{\rm rec},
\label{eq:eta}
\end{equation}
the mean number of events in a deposition volume during one trapped-charge lifetime. With $R_{\rm loc}$ = 0.5--\SI{2}{\micro m} and $\tau_{\rm rec}$ = 1--\SI{60}{s}, $\eta$ at $10^{10}$\,cm$^{-2}$s$^{-1}$ is 12 (0.6--220) and at $10^9$ it is 1.2 (0.06--22); $\eta\ge1$ in 93\% of the parameter draws at $10^{10}$, 55\% at $10^9$ and 12\% at $10^8$. These fractions describe the parameter draws, not the probability that the physical system is in one regime or the other, and $\eta\ge1$ means only that events overlap, not that the continuous-source approximation has been shown to be accurate. On this criterion, $10^{10}$ lies in the overlapping regime for most of the admissible parameter space, $10^9$ is in a transition region, and $10^8$ and below are in the sparse regime.

\begin{figure}[t]
\centering
\includegraphics[width=0.7\textwidth]{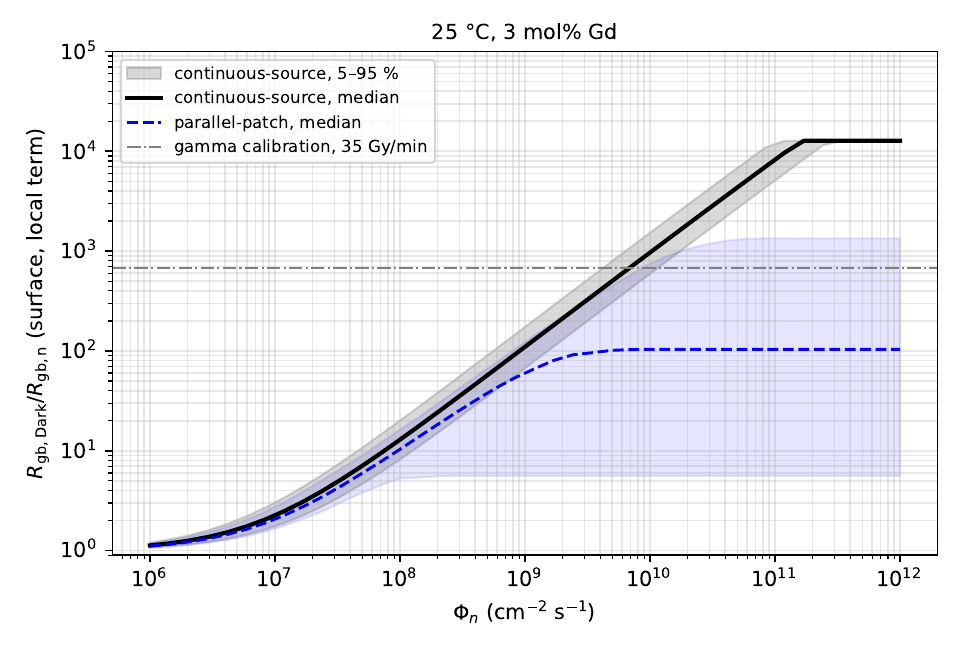}
\caption{Predicted surface-GB resistance ratio versus thermal-neutron flux at \SI{25}{\celsius}, \SI{3}{mol\percent} Gd, local term only. Black: continuous-source (mean-field) prediction, median and 5--95\% band from joint propagation of the parameter intervals. Blue: the parallel-patch estimate, median and band; the lower bound in the sparse regime is unity. The horizontal line is the gamma calibration value.}
\label{fig:RT}
\end{figure}

In the sparse regime, the response can be bounded but not predicted. An isolated capture delivers $N_{\rm pair}\approx\num{7e3}$ carriers to $V_{\rm loc}$ at once, of the same order as the number that the \SI{35}{Gy/min} calibration delivers to the same volume in one second. Two cases are considered. If the ionic lattice follows the trapped charge of each event ($\tau_{\rm ion}\lesssim\tau_{\rm rec}$, as the recovery data suggest), each affected GB patch takes the calibrated closed-form response to $G_{\rm local} = N_{\rm pair}f_{\rm enh}/(V_{\rm loc}\tau_{\rm rec})$ for a time $\tau_{\rm rec}$. A fraction $f_{\rm on} = 1-e^{-\eta}$ of the GB area is affected at any instant, and the patches act in parallel, giving $R_{\rm D}/\langle R\rangle = 1+f_{\rm on}(R_{\rm on}-1)$. This parallel-patch estimate is close to the mean-field value below $\eta = 1$ (Fig.~\ref{fig:RT}) because the logarithmic form of Eq.~\eqref{eq:seager} makes the time-averaged and the patch-averaged responses nearly equal; above $\eta = 1$ it falls below the mean-field value because $R_{\rm on}$ saturates. If instead the ions cannot follow individual events ($\tau_{\rm ion}\gg\tau_{\rm rec}$), no steady-state change results and the ratio is unity. Therefore, for $\eta<1$ the prediction is an interval $[1, R_{\rm MF}]$ with the upper end the more likely value; for $\eta\gg1$ the mean-field value is the prediction; and in the transition region around $\eta\approx1$ the mean-field value is an estimate whose error has not been quantified here, since neither the patch model nor the mean-field model is exact there. The room-temperature recovery time constant has only been bounded, and it determines which case applies at $10^9$\,cm$^{-2}$s$^{-1}$.

\section{Sample-level response}
\label{sec:sample}

\subsection{Self-shielding}
With $\Sigma_a = \SI{36.7}{cm^{-1}}$ at \SI{3}{mol\percent}, the thermal-neutron mean free path is \SI{0.27}{mm}; the \SI{0.8}{mm} pellet of \cite{defferriere2024ionic} absorbs 95\% of a normally incident beam. The GB resistance of the pellet is a series combination of layers at different local flux. For a one-directional beam, $\Phi(z) = \Phi_0e^{-\Sigma_az}$ (no scattering; an isotropic field gives a milder dependence), $N$ slabs of equal dark GB resistance give a pellet-level ratio $N/\sum_i(\text{ratio}_i)^{-1}$. For the local term alone, at $\Phi_0 = 10^{10}$\,cm$^{-2}$s$^{-1}$ and \SI{25}{\celsius} the surface-GB value of 970 (600--1500) becomes 550 (340--870) for \SI{0.3}{mm}, 180 (110--280) for \SI{0.8}{mm} and 9 (7--13) for \SI{2}{mm}: the deeper layers remain nearly dark and dominate the series resistance.

\subsection{X-ray, capture-photon and background-gamma terms}
\begin{figure}[t]
\centering
\includegraphics[width=\textwidth]{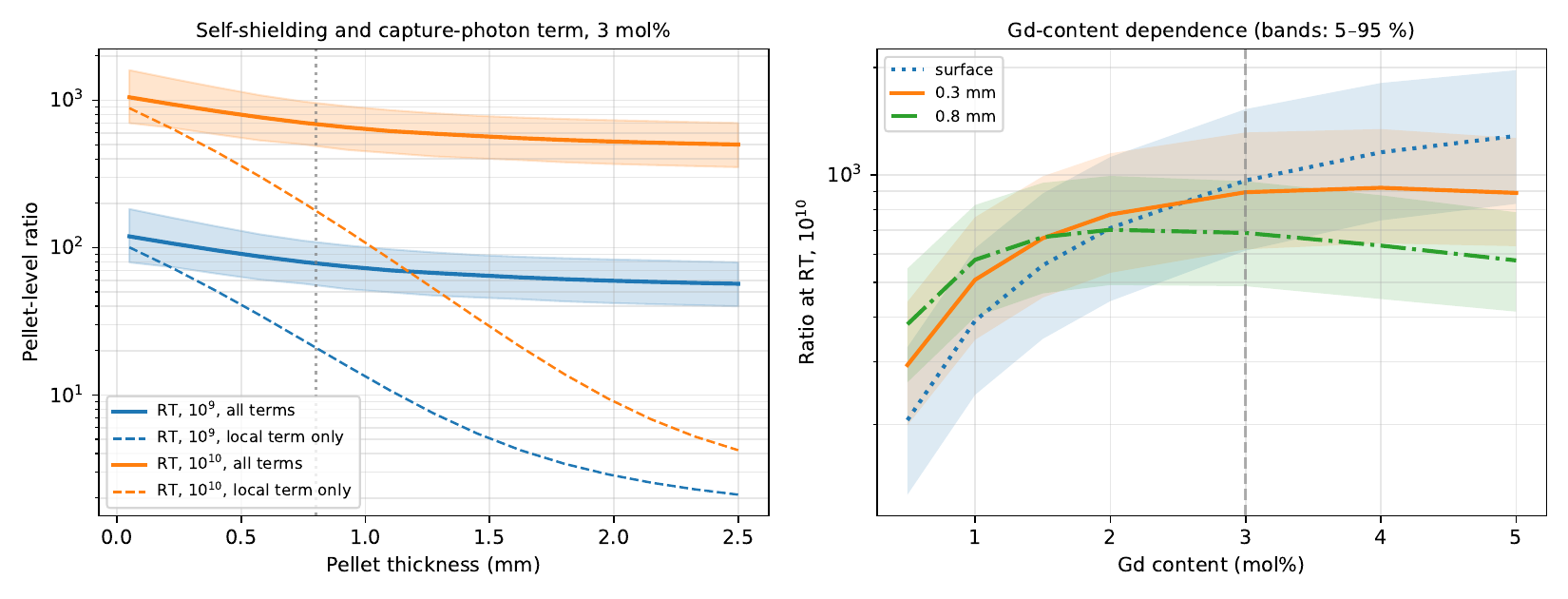}
\caption{(a) Pellet-level resistance ratio versus thickness at \SI{25}{\celsius} for $\Phi_0 = 10^9$ and $10^{10}$\,cm$^{-2}$s$^{-1}$, one-directional beam, \SI{3}{mol\percent} Gd: solid lines and bands, median and 5--95\% range including self-shielding and the capture-photon term; dashed, median of the local term alone; the dotted line marks the \SI{0.8}{mm} pellet of \cite{defferriere2024ionic}. (b) Ratio versus Gd content at $10^{10}$\,cm$^{-2}$s$^{-1}$ for the surface GB and for 0.3 and \SI{0.8}{mm} pellets, median and band.}
\label{fig:pellet}
\end{figure}

Two parts of the capture energy are deposited in the pellet but not within $R_{\rm loc}$ of the capture site: the Gd X-rays (\SI{11}{keV} per capture, absorbed within about \SI{0.1}{mm}) and the Compton-absorbed fraction of the cascade photons. With $\mu_{\rm en}/\rho$ = 0.019--\SI{0.029}{cm^2/g} for 1--3 MeV photons, the latter is $1-e^{-\mu_{\rm en}L}$ of about \SI{7.9}{MeV}, i.e.\ about 40, 110 and \SI{270}{keV} per capture for 0.3, 0.8 and \SI{2}{mm}. Together they enter Eq.~\eqref{eq:seager} as a uniform generation term equal to 0.6 (0.4--0.9) and 1.4 (0.9--2.1) times the GB-effective local term for 0.3 and \SI{0.8}{mm}. This term is physically a gamma response of the pellet to its own capture photons. It partly compensates the self-shielding of the local term, because it increases with thickness, and it is included in all pellet-level values quoted in this paper. Including both, at $\Phi_0 = 10^{10}$\,cm$^{-2}$s$^{-1}$ and \SI{25}{\celsius} the pellet-level ratio is 1030 (690--1540) for \SI{0.1}{mm}, 910 (630--1310) for \SI{0.3}{mm}, 700 (500--940) for \SI{0.8}{mm} and 530 (380--710) for \SI{2}{mm} (Fig.~\ref{fig:pellet}a); for pellets thicker than about \SI{0.5}{mm} most of this response is due to the capture photons. Because $\Sigma_a\propto x_{\rm Gd}$, raising the Gd content thins the active layer: the optimum composition moves from about \SI{5}{mol\percent} for the surface to about 3--4 and \SI{2}{mol\percent} for 0.3 and \SI{0.8}{mm} pellets (Fig.~\ref{fig:pellet}b and Sec.~\ref{sec:gd}). A gamma background of the source adds a further uniform term; keeping it below 10\% of $G_{\rm eff,GB}$ requires a background dose rate below about \SI{0.04}{Gy/min} at $\Phi_n = 10^8$, \SI{0.4}{Gy/min} at $10^9$ and \SI{4}{Gy/min} at $10^{10}$\,cm$^{-2}$s$^{-1}$.

\section{Results}
\label{sec:results}

\subsection{Room-temperature predictions}
Table~\ref{tab:ref} lists the predicted surface-GB ratio at \SI{25}{\celsius} for the local term alone, as median and 5--95\% range from the joint propagation of Sec.~\ref{sec:uncert}; the uniform term of Sec.~\ref{sec:sample} adds a comparable contribution for real pellets. At $10^{10}$\,cm$^{-2}$s$^{-1}$ the continuous-source prediction is 970 (600--1500); $\eta\ge1$ in 93\% of the draws, which supports the use of the continuous-source form but does not quantify its error. At $10^9$ the continuous-source value is 110 (70--170), but $\eta\ge1$ in only 55\% of the draws, so this value is a transition-region estimate whose model error is not quantified. At $10^8$ the continuous-source value is 13 (8--20), $\eta<1$ in 88\% of the draws, and the prediction is the interval $[1,13]$. The flux dependence in the continuous regime is close to one decade of ratio per decade of flux, the signature of the logarithmic Eq.~\eqref{eq:seager} exponentiated by Eq.~\eqref{eq:VW}.

\begin{table}[t]
\centering
\caption{Predicted GB resistance ratio at \SI{25}{\celsius}, surface of a \SI{3}{mol\percent} sample, local term only: median and 5--95\% range from joint propagation of the parameter intervals of Table~\ref{tab:params}. MF: continuous-source value; patch: parallel-patch estimate; $P(\eta\ge1)$: fraction of draws in the continuous regime.}
\label{tab:ref}
\begin{tabular}{l l l l l}
\toprule
$\Phi_n$ (cm$^{-2}$s$^{-1}$) & MF & patch & $\eta$ & $P(\eta\ge1)$ \\
\midrule
$10^7$  & 2.3 (1.8--3.0) & 2.1 (1.7--2.7) & 0.01 (0.0006--0.2) & 0.00 \\
$10^8$  & 13 (8--20) & 10 (6--17) & 0.12 (0.006--2.2) & 0.12 \\
$10^9$  & 110 (70--170) & 59 (7--120) & 1.2 (0.06--22) & 0.55 \\
$10^{10}$ & 970 (600--1500) & 92 (7--730) & 12 (0.6--220) & 0.93 \\
$10^{11}$ & 8300 (5200--12800) & 92 (7--1450) & 120 (6--2200) & 1.00 \\
\bottomrule
\end{tabular}
\end{table}

\subsection{Temperature window}
\begin{figure}[t]
\centering
\includegraphics[width=\textwidth]{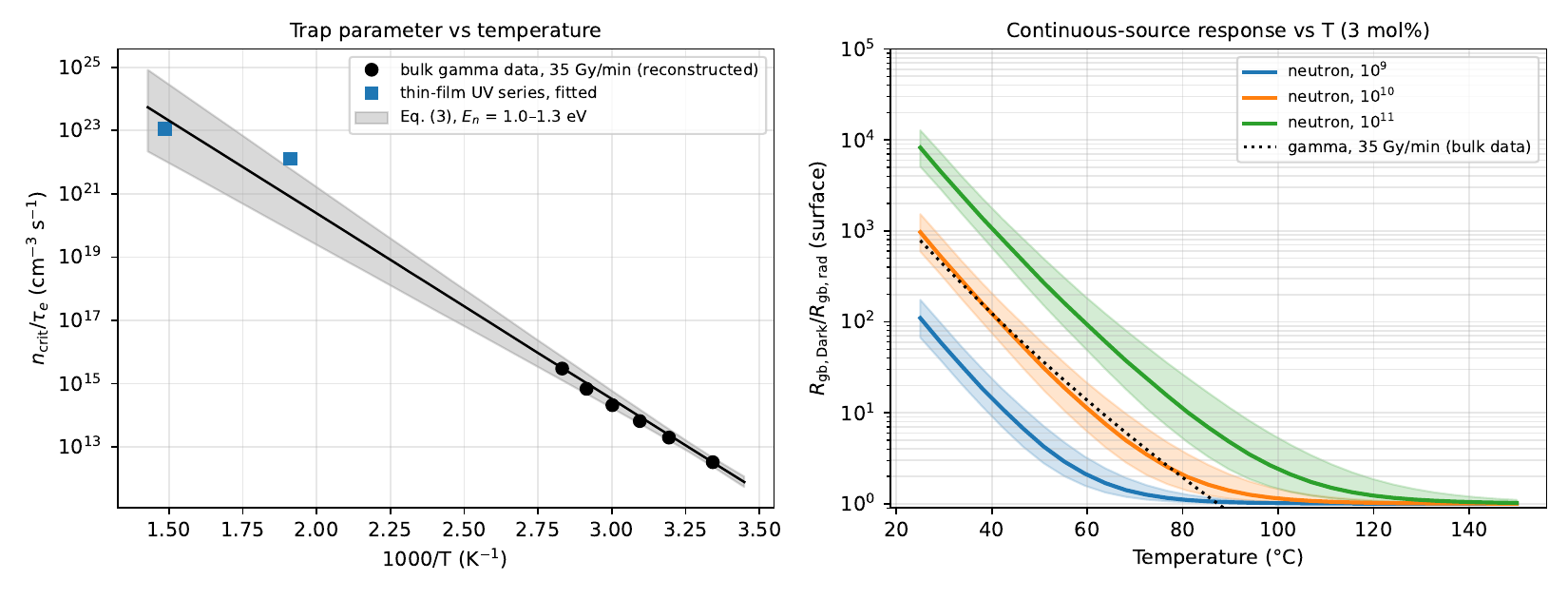}
\caption{(a) Trap parameter $n_{\rm crit}/\tau_e$ versus inverse temperature: circles, inverted from the reconstructed gamma ratio(T) of \cite{defferriere2024ionic} (these follow the stated activation energies by construction); squares, fitted independently to the thin-film UV intensity series of \cite{defferriere2022photo} at 250 and \SI{400}{\celsius}; line and band, Eq.~\eqref{eq:ncT} with $E_n$ = 1.0--\SI{1.3}{eV} and the calibration interval. (b) Continuous-source resistance ratio (local term, surface GB) versus temperature at three thermal fluxes, median and 5--95\% band, compared with the reconstructed gamma response at \SI{35}{Gy/min}.}
\label{fig:T}
\end{figure}

With Eq.~\eqref{eq:ncT}, the neutron response falls with temperature exactly as the gamma response does at the same $G_{\rm eff,GB}$ (Fig.~\ref{fig:T}b). For the surface GB at $10^9$\,cm$^{-2}$s$^{-1}$ the ratio drops from 110 (70--170) at \SI{25}{\celsius} to 15 (9--24) at \SI{40}{\celsius} and reaches 2 at 61 (55--68)\,$^\circ$C; at $10^{10}$ it is 130 (80--210) at \SI{40}{\celsius}, 12 (6--22) at \SI{60}{\celsius} and reaches 2 at 81 (73--92)\,$^\circ$C; at $10^{11}$ it reaches 2 at 105 (93--119)\,$^\circ$C. For a \SI{0.3}{mm} pellet including the capture-photon term the values are somewhat higher. Each decade of flux buys roughly \SI{20}{\celsius} of operating window. This is a direct consequence of the \SI{1.1}{eV} activation of the trap parameter. This is the main constraint provided by the gamma temperature series. A neutron experiment should therefore be performed as close to room temperature as the sample stage allows.

\subsection{Gd content}
\label{sec:gd}
Four factors change with $x_{\rm Gd}$: the capture rate ($\propto x$), the self-shielding ($\Sigma_a\propto x$), the dark barrier $\Delta\Phi_{\rm D}(x)$ and, through it, the ceiling $R_{\rm gb,D}/R_{\rm bulk}$. The trap parameter $n_{\rm crit}$ is calibrated at \SI{3}{mol\percent} only and is held independent of $x$; the exponent of $\Delta\Phi_{\rm D}(x)$ is carried as 0.45--0.65. Table~\ref{tab:gd} gives the room-temperature predictions for the compositions of a typical series, for the gamma calibration condition as well as for neutrons, so that the composition dependence can be tested under gamma irradiation before any neutron experiment.

\begin{table}[t]
\centering
\caption{Room-temperature GB resistance ratio versus Gd content, median and 5--95\% range from joint propagation: the gamma calibration condition, the neutron surface-GB value at $10^9$ and $10^{10}$\,cm$^{-2}$s$^{-1}$ (local term), and the \SI{0.3}{mm} pellet-level value at $10^{10}$ (self-shielding and capture-photon term included). The ceiling $R_{\rm gb,D}/R_{\rm bulk}$ is interpolated between the 3 and \SI{20}{mol\percent} anchors. $^\ddagger$: not given, because the required $\Delta\Phi_{\rm L}$ lies below the validity range of Eq.~\eqref{eq:VW} ($ze\Delta\Phi_{\rm L}/k_{\rm B}T<1$).}
\label{tab:gd}
\footnotesize
\begin{tabular}{@{}l l l l l l l@{}}
\toprule
$x$ (mol\%) & $\Delta\Phi_{\rm D}$ (V) & $\gamma$, 35 Gy/min & n, $10^9$ & n, $10^{10}$ & ceiling & 0.3 mm pellet, $10^{10}$ \\
\midrule
1  & 0.37 (0.33--0.40) & 810 (600--1030) & 42 (26--67) & 390 (240--630) & 2.0e4 & 520 (350--760) \\
2  & 0.25 (0.24--0.26) & 750 (560--950) & 78 (49--124) & 720 (440--1130) & 1.6e4 & 790 (540--1150) \\
3  & 0.20 & 690 (520--880) & 110 (70--170) & 970 (600--1500) & 1.3e4 & 910 (630--1310) \\
5  & 0.15 (0.14--0.16) & 600 (450--750) & 160 (100--250) & 1290 (830--1960) & 8.2e3 & 900 (650--1260) \\
10 & 0.10 (0.09--0.12) & 390 (290--500) & 210 (140--320) & n/a$^\ddagger$ & 2.8e3 & n/a$^\ddagger$ \\
20 & 0.07 (0.06--0.08) & n/a$^\ddagger$ & n/a$^\ddagger$ & n/a$^\ddagger$ & 3.1e2 & n/a$^\ddagger$ \\
\bottomrule
\end{tabular}
\end{table}

Between 1 and \SI{5}{mol\percent} the response at a given flux rises with $x$ because the capture rate rises faster than the barrier falls; the surface-GB optimum is near \SI{5}{mol\percent} and the pellet-level optimum near 3--\SI{4}{mol\percent} for sub-mean-free-path samples (Sec.~\ref{sec:sample}). Above \SI{5}{mol\percent} the limiting factor is the dark barrier itself: with $e\Delta\Phi_{\rm D}$ at about 4 and 3\,$k_{\rm B}T$ for 10 and \SI{20}{mol\percent}, the barrier is only a few $k_{\rm B}T$ high in the dark and leaves little room for a large ratio; Eq.~\eqref{eq:VW} cannot be used to quantify the remaining range. The $\Delta\Phi_{\rm D}(x)$ relation is an extrapolation outside \SI{3}{mol\percent}, so the rows above \SI{5}{mol\percent} are indicative only; for 10 and \SI{20}{mol\percent} the predicted $\Delta\Phi_{\rm L}$ at $10^{10}$\,cm$^{-2}$s$^{-1}$ falls below the validity range of Eq.~\eqref{eq:VW} and no value is given. The capture rate and the barrier can also be decoupled: selective doping of the GB core with Al raises $\Delta\Phi_{\rm D}$ of 3GDC to about \SI{0.30}{V} and with Sr lowers it to about \SI{0.13}{V} while the bulk Gd content and bulk conductivity are unchanged \cite{sha2026chemical}, which would allow the dark barrier to be set independently of the Gd content that fixes $R_{\rm cap}$ and $\Sigma_a$. Within the assumed composition model, the ordering in the table does not depend on these extrapolations. In particular, a composition series measured under gamma irradiation at \SI{35}{Gy/min} (of order $10^3$ at 1--3\,mol\%, a few hundred at 5--10, tens at 20) can test the $x$-dependence of the framework without neutron beam time.

\subsection{Propagation of the parameter intervals}
\label{sec:uncert}
All interval parameters of Table~\ref{tab:params} are sampled jointly (2000 draws): $f_{\rm enh}\sim U[1,2]$; local energy per capture $U[49,75]$\,keV for $^{157}$Gd and $U[25,75]$\,keV for $^{155}$Gd; $R_{\rm loc}$ log-normal with 90\% range 0.5--\SI{2}{\micro m}; $W_{\rm pair}\sim U[7.6,10.2]$\,eV; calibration ratio $\sim U[500,900]$; $E_n\sim U[1.0,1.3]$\,eV; $\tau_{\rm rec}$ log-uniform on 1--\SI{60}{s}; $\mu_{\rm en}/\rho\sim U[0.019,0.029]$\,cm$^2$/g; and the exponent of $\Delta\Phi_{\rm D}(x)$ $\sim U[0.45,0.65]$. $n_{\rm crit}/\tau_e$ is re-fitted for each draw, and every quantity in the paper is the median and 5--95\% range of the resulting distribution. At \SI{25}{\celsius} the width of the continuous-source interval is set mainly by $f_{\rm enh}$ and the calibration ratio; at elevated temperature by $E_n$; and in the regime classification by $R_{\rm loc}$ and $\tau_{\rm rec}$. These intervals propagate the stated parameter ranges through the model as written. They do not include the structural uncertainties of the model itself: the mean-field treatment of a distribution of GB potentials, the asymptotic form of Eq.~\eqref{eq:seager}, the transition-region error at $\eta\approx1$, the one-directional beam geometry, and the assumption that $n_{\rm crit}$ and the local deposition are the same for all compositions and for both Gd isotopes. The intervals are therefore conditional parameter-propagation intervals and lower bounds on the prediction uncertainty.

\section{Independent check: thin-film UV data}
\label{sec:transfer}

\begin{figure}[t]
\centering
\includegraphics[width=0.6\textwidth]{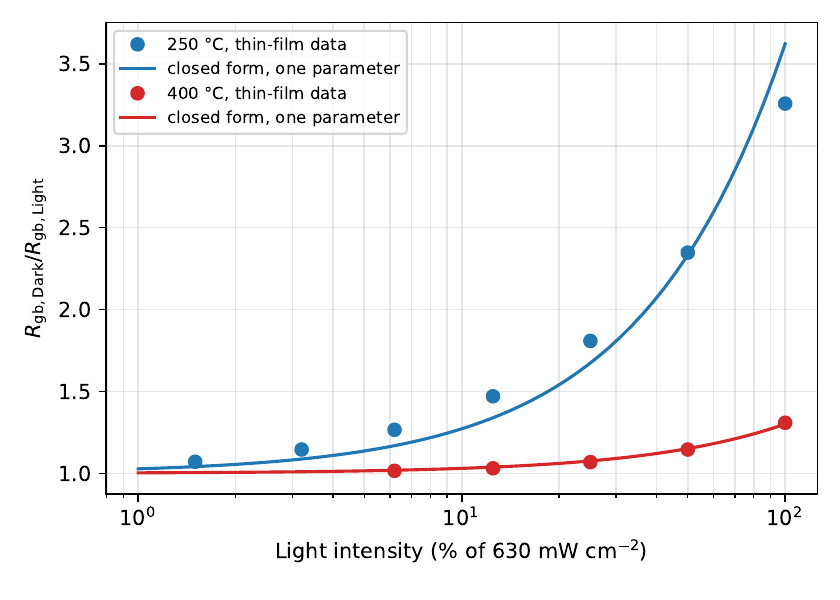}
\caption{Thin-film GB resistance ratio versus UV intensity at 250 and \SI{400}{\celsius}, Supplementary Table~S3 of \cite{defferriere2022photo}, compared with Eqs.~\eqref{eq:VW}--\eqref{eq:seager} with $n\propto$ intensity and one fitted $n_{\rm crit}/\tau_e$ per temperature.}
\label{fig:s3}
\end{figure}

The thin-film measurements of \cite{defferriere2022photo} were not used in the calibration, so they provide an independent check. Their Supplementary Information gives, for one film, a seven-point intensity series at \SI{250}{\celsius} and a five-point series at \SI{400}{\celsius} (Table~S3) and the generation rate at full intensity, $G = \SI{3.7e22}{cm^{-3}s^{-1}}$ (Section~17), with $\Delta\Phi_{\rm D} = \SI{0.25}{V}$. With one fitted $n_{\rm crit}/\tau_e$ per temperature, Eqs.~\eqref{eq:VW}--\eqref{eq:seager} reproduce the \SI{400}{\celsius} series to 1\% and the \SI{250}{\celsius} series to 7\% (r.m.s.\ in $\ln$ ratio) over a 70-fold intensity range (Fig.~\ref{fig:s3}). This indicates that the functional form, and hence the predicted flux dependence, is consistent with the only intensity-resolved data available. The fitted values are \SI{1.3e22}{cm^{-3}s^{-1}} at \SI{250}{\celsius} and \SI{1.2e23}{} at \SI{400}{\celsius}. Equation~\eqref{eq:ncT}, calibrated entirely on the bulk gamma data between 26 and \SI{80}{\celsius} and extrapolated by 375 degrees, predicts 1.9 (0.1--34)$\times10^{23}$ at \SI{400}{\celsius}, which brackets the fitted value, and 6.5 (0.7--59)$\times10^{20}$ at \SI{250}{\celsius}, 20 (2--200) times below the fitted value. The \SI{250}{\celsius} discrepancy is in the direction expected from the constriction picture of the original authors, in which at low temperature only a few percent of parallel pathways collapse and a mean-field $\Delta\Phi$ overstates the response (their Section~14). The two data sets differ in sample form (\SI{250}{nm} film versus \SI{0.8}{mm} pellet), grain size (50--100 versus \SI{400}{nm}) and excitation. The two data sets therefore agree within a factor of order 10 over ten orders of magnitude of $n_{\rm crit}/\tau_e$, which is consistent with the framework; the \SI{250}{\celsius} offset shows that the bulk calibration does not transfer to the thin film quantitatively.

\section{Discrimination and confounders}
\label{sec:confounders}

A decrease of GB impedance under neutron irradiation must be attributed to the capture-localized radiation-ionic mechanism rather than to the alternatives listed below. The proposed experiment in Sec.~\ref{sec:protocol} is designed to separate them.

\emph{Gamma-mediated response.} Two gamma contributions are present: the external background of the source and the capture photons generated in the pellet itself (Sec.~\ref{sec:sample}). A cadmium-shielded exposure at the same position removes the thermal neutrons and therefore removes both the local electron term and the capture-photon term; what remains is the response to the external gamma background and to any fast neutrons. The difference between the unshielded and the Cd-shielded exposure is therefore the total capture-induced response (local electrons plus capture photons). The two capture-induced terms are not separated by the Cd control, nor by a Gd-free reference pellet exposed on its own; their different thickness dependence (Fig.~\ref{fig:pellet}a), or a Gd-free pellet placed in contact with the Gd pellet so that it receives the capture photons, constrains their relative size without separating them uniquely. Because the response is non-linear in the generation rate, the capture-induced response is not obtained by subtracting the Cd-shielded resistance ratio from the unshielded one; the two exposures should instead be compared by inserting their respective source terms into the model.

\emph{Electronic conduction.} In the gamma experiment the bulk arc was unchanged to within 1\% and its capacitance unchanged, which excludes a significant electronic contribution \cite{defferriere2024ionic}; the same test applies, and is sharper for neutrons because the Gd-free reference carries any electronic response of the gamma field without the capture term.

\emph{Heating.} At $10^{10}$\,cm$^{-2}$s$^{-1}$ the capture rate of $\SI{3.7e11}{cm^{-3}s^{-1}}$ releases \SI{0.47}{W/cm^3}, almost all of which leaves the pellet as photons. The power absorbed in the pellet, averaged over its thickness, is the local term plus the X-ray and Compton terms of Sec.~\ref{sec:sample}: about \SI{6}{mW/cm^3} for \SI{0.3}{mm} and \SI{10}{mW/cm^3} for \SI{0.8}{mm}. This is comparable to the \SI{4}{mW/cm^3} absorbed in the gamma experiment (\SI{35}{Gy/min} in CeO$_2$), for which a temperature rise of \SI{0.1}{K} was inferred \cite{defferriere2024ionic}.

\emph{Displacement damage.} The $^{158}$Gd recoil from a cascade photon of energy $E_\gamma$ is $E_\gamma^2/2Mc^2$, \SI{14}{eV} at \SI{2}{MeV} and \SI{210}{eV} for the full \SI{7.9}{MeV} in a single photon; only the rare high-energy lines exceed the threshold displacement energies of ceria. Even at one displacement per capture the rate at $10^{10}$\,cm$^{-2}$s$^{-1}$ is $\sim10^{-4}$\,dpa per year. The gamma experiment found no structural or Raman change after \SI{1}{MGy} \cite{defferriere2024ionic}. Fast neutrons in the field, if present, are removed by the Gd-free reference.

\section{Proposed falsification experiment}
\label{sec:protocol}

The prediction can be tested with the following source and sample parameters.
\begin{enumerate}
\item \emph{Sample.} Polycrystalline 3GDC pellets of two thicknesses, about 0.1 and \SI{0.3}{mm}, prepared as in \cite{defferriere2024ionic} (\SI{400}{nm} grains, blocking Pt electrodes), with the GB arc resolved in the dark. The thickness pair constrains the relative size of the local electron term and the capture-photon term (Sec.~\ref{sec:sample}). A composition series (1, 3, 5, \SI{10}{mol\percent}) characterized first under gamma irradiation tests Table~\ref{tab:gd} without neutrons.
\item \emph{Temperature.} As close to \SI{25}{\celsius} as possible; each \SI{20}{\celsius} costs a decade of response (Fig.~\ref{fig:T}b).
\item \emph{Flux.} Thermal flux of $10^{10}$\,cm$^{-2}$s$^{-1}$ at the sample, uniform over the electrode area to a factor of two, with a second value at $10^{9}$ to test the predicted slope and the transition region.
\item \emph{Gamma background.} Below about \SI{0.4}{Gy/min} at $10^9$ and \SI{4}{Gy/min} at $10^{10}$\,cm$^{-2}$s$^{-1}$; a Cd-shielded exposure at the same position measures the response to the external background and to fast neutrons.
\item \emph{Transients.} Record the impedance at one frequency through beam-on and beam-off. The beam-off recovery gives $\tau_{\rm rec}$ at room temperature, which the gamma work bounds only as ``seconds'' and which decides whether $10^9$\,cm$^{-2}$s$^{-1}$ is in the continuous regime (Sec.~\ref{sec:regimes}).
\item \emph{Predicted outcome.} For a \SI{0.3}{mm} pellet at $10^{10}$\,cm$^{-2}$s$^{-1}$ and \SI{25}{\celsius}, a GB resistance decrease by a factor of 910 (630--1310), of which about 40\% is due to the capture photons; for the \SI{0.1}{mm} pellet 1030 (690--1540), of which about 20\% is due to the capture photons. At $10^9$ the corresponding value is 100 (70--150) if the continuous description holds, with unity admitted if the recovery time proves shorter than about \SI{10}{s}. These intervals are conditional on the model and on the parameter distributions of Sec.~\ref{sec:uncert}, and no numerical falsification threshold is derived here, because the structural error of the model (the transition-region behaviour, the mean-field treatment of the GB distribution and the trapped-charge kinetics) has not been quantified. Once the dose, temperature, microstructure and recovery time are confirmed to satisfy the assumptions of the model, a measured decrease well below the predicted interval challenges the model; the comparison should be made by inserting the measured source terms of the unshielded and Cd-shielded exposures into the model rather than by subtracting resistance ratios.
\end{enumerate}

\section{Conclusions}

In this work, the response of the grain-boundary impedance of 3GDC to thermal neutron irradiation was modelled using the radiation-ionic framework calibrated on the published gamma data. The main results are as follows.

\begin{enumerate}
\item In the continuous-source limit, a thermal flux of 8 (6--12)$\times10^{9}$\,cm$^{-2}$s$^{-1}$ delivers to the grain boundaries the same generation rate as the \SI{35}{Gy/min} gamma calibration. The predicted surface-GB resistance ratio at room temperature is 110 (70--170) at $10^9$ and 970 (600--1500) at $10^{10}$\,cm$^{-2}$s$^{-1}$, with roughly one decade of ratio per decade of flux. The locally deposited energy per capture, 49--\SI{75}{keV} in conversion and Auger electrons, is taken from the gadolinium neutron capture therapy literature and sets the scale of the prediction. Because the conversion-electron range exceeds the grain size, the segregation of Gd to the grain boundaries increases the response by a factor between 1 and 2, and this interval is the largest single contribution to the width of the room-temperature prediction.
\item The gamma temperature series fixes the temperature dependence of the trap parameter at $E_n$ = 1.0--\SI{1.3}{eV}. This range is consistent with hole-transport activation, brackets the trap parameter fitted independently to the thin-film UV data at \SI{400}{\celsius}, and is 2--200 times below it at \SI{250}{\celsius}. The neutron response falls to 2 at 61 (55--68)\,$^\circ$C for $10^9$ and 81 (73--92)\,$^\circ$C for $10^{10}$\,cm$^{-2}$s$^{-1}$.
\item Capture events are discrete. The continuous-source description holds above a critical flux set by the room-temperature trapped-charge lifetime, which the gamma experiment bounds only loosely. At $10^{10}$\,cm$^{-2}$s$^{-1}$, 93\% of the parameter draws satisfy $\eta\ge1$, and at $10^9$ 55\%, although this criterion alone does not establish the accuracy of the continuous-source approximation. The beam-off recovery transient is therefore the first quantity that a neutron experiment should record.
\item The capture photons absorbed in the pellet produce a uniform gamma response equal to 0.4--0.9 of the local electron response for a \SI{0.3}{mm} pellet and 0.9--2.1 for \SI{0.8}{mm}. A large part of the neutron response of a millimetre pellet is therefore due to its own capture photons, and their relative size is constrained by the thickness dependence. Within the assumed composition model, the response above \SI{5}{mol\percent} Gd is limited by the weak dark barrier at any flux.
\end{enumerate}

The experiment required to test these predictions is specified in terms of flux, gamma background, sample thickness and controls. A composition series measured under gamma irradiation can test the composition dependence of the framework before any neutron experiment is performed. This work will aid the design of neutron irradiation experiments on doped ceria and related solid electrolytes, and the evaluation of ionic-conduction-based materials for neutron detection.

\section*{Acknowledgements}
The author thanks Prof.\ Wei Zheng (Lanzhou University) and Prof.\ Colin Boxall (Lancaster University) for discussions, and Dr Andrew Parker (Lancaster University) for his help with earlier funding applications related to this work.

\bibliographystyle{unsrt}
\bibliography{ref}

\begin{thebibliography}{10}

\bibitem{avila2009dopant}
Hugo~J Avila-Paredes, Kwanghoon Choi, Chien-Ting Chen, and Sangtae Kim.
\newblock Dopant-concentration dependence of grain-boundary conductivity in
  ceria: A space-charge analysis.
\newblock {\em Journal of Materials Chemistry}, 19(27):4837--4842, 2009.

\bibitem{kim2016determining}
Sangtae Kim, Seong~K Kim, Sergey Khodorov, Joachim Maier, and Igor Lubomirsky.
\newblock On determining the height of the potential barrier at grain
  boundaries in ion-conducting oxides.
\newblock {\em Physical Chemistry Chemical Physics}, 18(4):3023--3031, 2016.

\bibitem{defferriere2022photo}
Thomas Defferriere, Dino Klotz, Juan~Carlos Gonzalez-Rosillo, Jennifer~LM Rupp,
  and Harry~L Tuller.
\newblock Photo-enhanced ionic conductivity across grain boundaries in
  polycrystalline ceramics.
\newblock {\em Nature Materials}, 21(4):438--444, 2022.

\bibitem{defferriere2024ionic}
Thomas Defferriere, Ahmed~Sami Helal, Ju~Li, Jennifer~LM Rupp, and Harry~L
  Tuller.
\newblock Ionic conduction-based polycrystalline oxide gamma ray
  detection--radiation-ionic effects.
\newblock {\em Advanced Materials}, 36(24):2309253, 2024.

\bibitem{defferriere2025optoionics}
Thomas Defferriere and Harry~L Tuller.
\newblock Optoionics: New opportunity for ionic conduction-based radiation
  detection.
\newblock {\em MRS communications}, 15(3):523--532, 2025.

\bibitem{sha2026chemical}
Z.~Sha, T.~Defferriere, C.~Gilgenbach, J.~M. LeBeau, I.~Lubomirsky, and H.~L.
  Tuller.
\newblock Chemical control of space charge barriers and grain boundary
  resistances in polycrystalline ionic conductors.
\newblock {\em Advanced Functional Materials}, 36:e77635, 2026.

\bibitem{vollman1994grain}
Markus Vollman and Rainer Waser.
\newblock Grain boundary defect chemistry of acceptor-doped titanates: space
  charge layer width.
\newblock {\em Journal of the American Ceramic Society}, 77(1):235--243, 1994.

\bibitem{bowman2015electrical}
William~J Bowman, Jiangtao Zhu, Renu Sharma, and Peter~A Crozier.
\newblock Electrical conductivity and grain boundary composition of gd-doped
  and gd/pr co-doped ceria.
\newblock {\em Solid State Ionics}, 272:9--17, 2015.

\bibitem{seager1981grain}
CH~Seager.
\newblock Grain boundary recombination: Theory and experiment in silicon.
\newblock {\em Journal of Applied Physics}, 52(6):3960--3968, 1981.

\bibitem{joshi1990theory}
Dinesh~Prasad Joshi and Devesh~Prasad Bhatt.
\newblock Theory of grain boundary recombination and carrier transport in
  polycrystalline silicon under optical illumination.
\newblock {\em IEEE transactions on electron devices}, 37(1):237--249, 1990.

\bibitem{klein1968bandgap}
Claude~A Klein.
\newblock Bandgap dependence and related features of radiation ionization
  energies in semiconductors.
\newblock {\em Journal of Applied Physics}, 39(4):2029--2038, 1968.

\bibitem{HELMER2004325}
R.G. Helmer.
\newblock Nuclear data sheets for a=158.
\newblock {\em Nuclear Data Sheets}, 101(3):325--519, 2004.

\bibitem{goorley2000electron}
Tim Goorley and Hooshang Nikjoo.
\newblock Electron and photon spectra for three gadolinium-based cancer therapy
  approaches.
\newblock {\em Radiation research}, 154(5):556--563, 2000.

\bibitem{berger2017nist}
MJ~Berger, JS~Coursey, MA~Zucker, and J~Chang.
\newblock Nist standard reference database 124.
\newblock {\em Last Update to Data Content: July}, 2017.

\end{thebibliography}

\end{document}